\pdfoutput=1

\documentclass[11pt]{article}
\RequirePackage[OT1]{fontenc}
\RequirePackage{amsthm,amsmath}
\RequirePackage[numbers]{natbib}
\RequirePackage[colorlinks,citecolor=blue,urlcolor=blue]{hyperref}

\usepackage[utf8]{inputenc} 
\usepackage[T1]{fontenc}    
\usepackage[colorlinks]{hyperref} 
\usepackage{url}            
\usepackage{booktabs}       
\usepackage{amsfonts}       
\usepackage{amsmath}
\usepackage{nicefrac}       
\usepackage{microtype}      
\usepackage{color}
\usepackage{bbm}
\usepackage{amssymb, enumerate}
\usepackage{makecell}
\usepackage{wrapfig}
\usepackage{extarrows}
\usepackage{arydshln}
\usepackage{titlesec}

\usepackage{caption}
\usepackage{multirow} 

\titleformat{\paragraph}
{\normalfont\normalsize\bfseries}{\theparagraph}{1em}{}
\titlespacing*{\paragraph}
{0pt}{3.25ex plus 1ex minus .2ex}{1.5ex plus .2ex}

\usepackage{amsmath,amsfonts,amsthm, amssymb, enumerate}
\usepackage{lipsum}
\usepackage{caption}
\usepackage{amssymb,graphicx,subfigure}
\usepackage[bottom]{footmisc}
\usepackage[dvipsnames]{xcolor}
\usepackage{array,multirow}
\usepackage{enumitem}

\usepackage{fancyhdr}
\usepackage[utf8]{inputenc}
\usepackage{float, nccmath, comment, amsmath}
\usepackage{multicol, multirow, booktabs, threeparttable, indentfirst, rotating}
\usepackage{graphics, graphicx}
\graphicspath{ {/} }

\usepackage{microtype}
\usepackage{graphicx}
\usepackage{subfigure}
\usepackage{booktabs} 
 
\usepackage{graphicx}
\usepackage{amsmath}
\usepackage{amssymb}
\usepackage{booktabs}  
\usepackage{times}
\usepackage{epsfig}
\usepackage{graphicx}
\usepackage{amsfonts, amsthm, enumerate} 
\usepackage[bottom]{footmisc} 
\usepackage{enumitem}
\usepackage{setspace}
\usepackage{wrapfig}
\usepackage{extarrows}
\usepackage{makecell}
\usepackage{bbm}  
\usepackage{textcomp}
\usepackage{xcolor} 
\usepackage{caption} 
\usepackage{kotex}

\usepackage{xr}
\makeatletter
\newcommand*{\addFileDependency}[1]{
  \typeout{(#1)}
  \@addtofilelist{#1}
  \IfFileExists{#1}{}{\typeout{No file #1.}}
}
\makeatother

\usepackage[font=small,labelfont=bf]{caption}

\usepackage{algorithm,algpseudocode}
\algnewcommand\TR{\item[{\textbf{Training phase}}]}
\algnewcommand\TE{\item[{\textbf{Test phase}}]}
\algnewcommand\Input{\item[{{Input:}}]}
\algnewcommand\Output{\item[{{Output:}}]}
\algnewcommand\Initialize{\item[{{Initialize:}}]}
\algnewcommand{\return}[1]{
	\State \textbf{return:}
	\Statex \hspace*{\algorithmicindent}\parbox[t]{.8\linewidth}{\raggedright #1}
}

 
\usepackage{setspace}
\usepackage[left=2cm,right=2cm,top=4cm,bottom=4cm,a4paper]{geometry}

\makeatletter
\newcommand\footnoteref[1]{\protected@xdef\@thefnmark{\ref{#1}}\@footnotemark}
\makeatother

\begin{document}

	\title{Comparative Validation of AI and non-AI Methods in MRI Volumetry to Diagnose Parkinsonian Syndromes} 
	\date{}
	\author{\parbox{\linewidth}{\centering
		Joomee Song, MD$^{1}$\thanks{Co-first authors, equally contributed to the work},
		Juyoung Hahm$^{2,3}$\footnotemark[1], 
		Jisoo Lee$^{2,4}$, 
		Chae Yeon Lim$^{2,5}$,
		Myung Jin Chung, MD, PhD$^{2,6,7}$, 
		Jinyoung Youn, MD, PhD$^{1}$,
		Jin Whan Cho, MD, PhD$^{1}$,
		Jong Hyeon Ahn, MD$^{1}$\thanks{Corresponding author (equally contributed as the co-corresponding author): Kyung-Su Kim (kskim.doc@gmail.com, Sungkyunkwan University School of Medicine, Seoul, Republic of Korea) and Jong Hyeon Ahn (jonghyeon.ahn@samsung.com, Department of Neurology, Samsung Medical Center, Sungkyunkwan University School of Medicine, Seoul, Republic of Korea)},\\
		Kyung-Su Kim, PhD$^{2,7}$\footnotemark[2]
		\\ 
		{\small $^{1}$Department of Neurology and Neuroscience Center, Samsung Medical Center, Sungkyunkwan University School of Medicine, Seoul, Republic of Korea}\\
		{\small $^{2}$Medical AI Research Center, Research Institute for Future Medicine, Samsung Medical Center, Seoul, Republic of Korea}\\
		{\small $^{3}$Department of Biostatistics, Columbia University, New York, NY, USA}\\
		{\small $^{4}$Department of Electrical and Computer Engineering, University of Maryland, College Park, MD, USA}\\
		{\small $^{5}$Department of Medical Device Management and Research, SAIHST, Sungkyunkwan University, Seoul, Republic of Korea}\\
		{\small $^{6}$Department of Radiology, Samsung Medical Center, Sungkyunkwan University School of Medicine, Seoul, Republic of Korea}\\
		{\small $^{7}$Department of Data Convergence and Future Medicine, Sungkyunkwan University School of Medicine, Seoul, Republic of Korea}
		}}
	\maketitle 
\begin{abstract}

\textbf{Background:} Automated segmentation and volumetry of brain magnetic resonance imaging (MRI) scans are essential for the diagnosis of Parkinson’s disease (PD) and Parkinson's plus syndromes (P-plus).

\textbf{Objective:} To enhance the diagnostic performance, we adopt deep learning (DL) models in brain segmentation and compared their performance with the gold-standard non-DL method.

\textbf{Methods:} We collected brain MRI scans of healthy controls ($n=105$) and patients with PD ($n=105$), multiple systemic atrophy ($n=132$), and progressive supranuclear palsy ($n=69$) at Samsung Medical Center from January 2017 to December 2020. Using the gold-standard non-DL model, FreeSurfer (FS), we segmented six brain structures: midbrain, pons, caudate, putamen, pallidum, and third ventricle, and considered them as annotating data for DL models, the representative V-Net and UNETR. The Dice scores and area under the curve (AUC) for differentiating normal, PD, and P-plus cases were calculated.

\textbf{Results:} The segmentation times of V-Net and UNETR for the six brain structures per patient were 3.48 ± 0.17 and 48.14 ± 0.97 s, respectively, being at least 300 times faster than FS (15,735 ± 1.07 s). Dice scores of both DL models were sufficiently high (>0.85), and their AUCs for disease classification were superior to that of FS. For classification of normal vs. P-plus and PD vs. multiple systemic atrophy (cerebellar type), the DL models and FS showed AUCs above 0.8.

\textbf{Conclusions:} DL significantly reduces the analysis time without compromising the performance of brain segmentation and differential diagnosis. Our findings may contribute to the adoption of DL brain MRI segmentation in clinical settings and advance brain research.

\end{abstract}

\section{Introduction}
Parkinson's disease (PD) diagnosis is primarily based on clinical presentation. However, for atypical symptoms called red flags \cite{berg2015mds}, brain magnetic resonance imaging (MRI) is essential for diagnosing Parkinson-plus syndromes (P-plus), such as multiple system atrophy (MSA) and progressive supranuclear palsy (PSP). MRI improves the diagnostic accuracy and can be used for monitoring disease progression \cite{meijer2017clinical}. Brain MRI can reveal various features that appear in P-plus but not in PD \cite{meijer2017clinical, watanabe2018clinical, whitwell2017radiological}. For instance, patients with PSP show marked midbrain atrophy \cite{jankovic2021principles}, known as the hummingbird sign. In MSA— Parkinsonian type (MSA-P), the putamen is atrophic, with a flattened lateral border, and shows a hypointense signal on T1-weighted gradient-echo images. Patients with MSA—cerebellar type (MSA-C) show predominant atrophy in the pons and middle cerebellar peduncles, resulting in an increased midbrain-to-pons ratio \cite{hussl2010diagnostic} and a decrease in the magnetic resonance Parkinsonism index \cite{quattrone2008mr}. Accordingly, quantitative measures of the volume of these brain structures have also been assessed, showing high sensitivity and specificity in differentiating PD from P-plus \cite{paviour2006regional}.

Although the diagnostic sensitivity and specificity obtained by evaluating the midbrain area are generally high for differentiating between PSP, MSA, and PD \cite{zanigni2016accuracy}, the visual assessment of this area is not quantitative, lacks objectivity, and highly dependent on the physician's skills or image acquisition. Consequently, diagnoses based on visual assessments have shown a broad spectrum of accuracy, even falling below 80\% \cite{massey2012conventional, schrag2000differentiation, kim2015visual}. To develop a consistent and quantitative analysis of brain MRI, volumetry of the midbrain area has been used as an optimal predictor for accurate diagnosis \cite{paviour2006regional, saeed2020neuroimaging, hussl2010diagnostic, moller2017manual}. Thus, brain image segmentation has become an important stage in most downstream analyses based on prediction models or automated machine-learning (ML) methods for volumetry and diagnosis.

A trained physician's manual segmentation of brain MRI scans is strenuous and time-consuming, and it requires a highly skilled specialist to correctly identify the brain structures. Various automated techniques using atlas-based or { deep-learning (DL)} techniques have been developed to overcome these problems. Although automated image segmentation models for the brain show limitations \cite {despotovic2015mri, fawzi2021brain},  FreeSurfer (FS) \cite{fischl2012freesurfer} can extract brain structures with relatively high accuracy. Therefore, FS has been widely adopted as a {non-DL} automated segmentation method \cite{fischl2012freesurfer, dewey_hana_2010, eggert_sommer_2012, mayer_latal_2016, klauschen_goldman_2009}.

Various automated segmentation methods for brain structures have been developed, but their use in clinical practice is limited, being typically used in one-time studies. This is attributable to the time-consuming and complex process of automated segmentation models compared with physicians' simple visual assessments of brain MRI scans.
For instance, the automated FS for segmentation takes more than 4.5 h per patient to segment the brain captured in an MRI scan. This complexity problem occurs because existing automated segmentation methods use atlas-based registration
\cite{pham2000current, christensen1997volumetric, collins1995automatic, iosifescu1997automated}. In fact, expressing segmentation as an atlas-based registration problem requires considerable time, and FS must be optimized to obtain a coordinate transformation function suitable for the internal atlas model of each test sample.

An automated model for fast segmentation and diagnosis without involving intricate methods should be developed for clinical use. Although DL segmentation has been used in various fields, including medicine \cite{fawzi2021brain}, the segmentation of brain structures in MRI for the diagnosis of neurodegenerative diseases has made little progress. In addition, no study has introduced artificial-intelligence-based analysis or demonstrated the usefulness of DL (i.e., complexity or disease discrimination performance) compared with existing non-DL automated segmentation of brain structures (e.g., FS). Unlike existing non-DL methods, DL may increase the analysis speed by completing segmentation using only forward computations based on learned parameters without requiring optimization processes such as registration. However, it is difficult to predict whether DL shows performance degradation compared with non-DL methods, especially in diagnosing neurodegenerative diseases. Our study is significant because it is the first experimental study that demonstrates, with extensive clinical data, the competitive performance of DL and non-DL methods. A DL method can achieve high performance in terms of the analysis complexity and diagnostic performance for differentiating major neurodegenerative diseases (e.g, differential diagnosis between PD, P-plus, and normal cases).

Recent DL segmentation models are classified into convolutional neural network (CNN) and vision transformer (ViT) architectures. {Accordingly, a representative model of each framework, V-Net \cite{milletari2016vnet} and UNet transformer (UNETR) \cite{hatamizadeh2022unetr}, respectively, were adopted to perform volumetric 3D image segmentation in this study.}
The DL {models were trained to segment brain structures on MRI scans for the diagnosis of} neurodegenerative diseases, and {their performances were} analyzed and compared with an existing non-DL model, FS. { Six brain structures that are important in classifying normal, PD, and P-plus cases were segmented: putamen, pallidum, midbrain, pons, caudate, and third ventricle. The volumes of the segmented areas were subsequently used to differentiate between normal, PD, and P-plus cases. We compared the disease differentiation accuracy and segmentation time of the DL models with those of FS, which were regarded as the reference (i.e., ground truth) for training the DL segmentation models.}

\begin{figure} 
\centering
\includegraphics[width=0.9\textwidth]{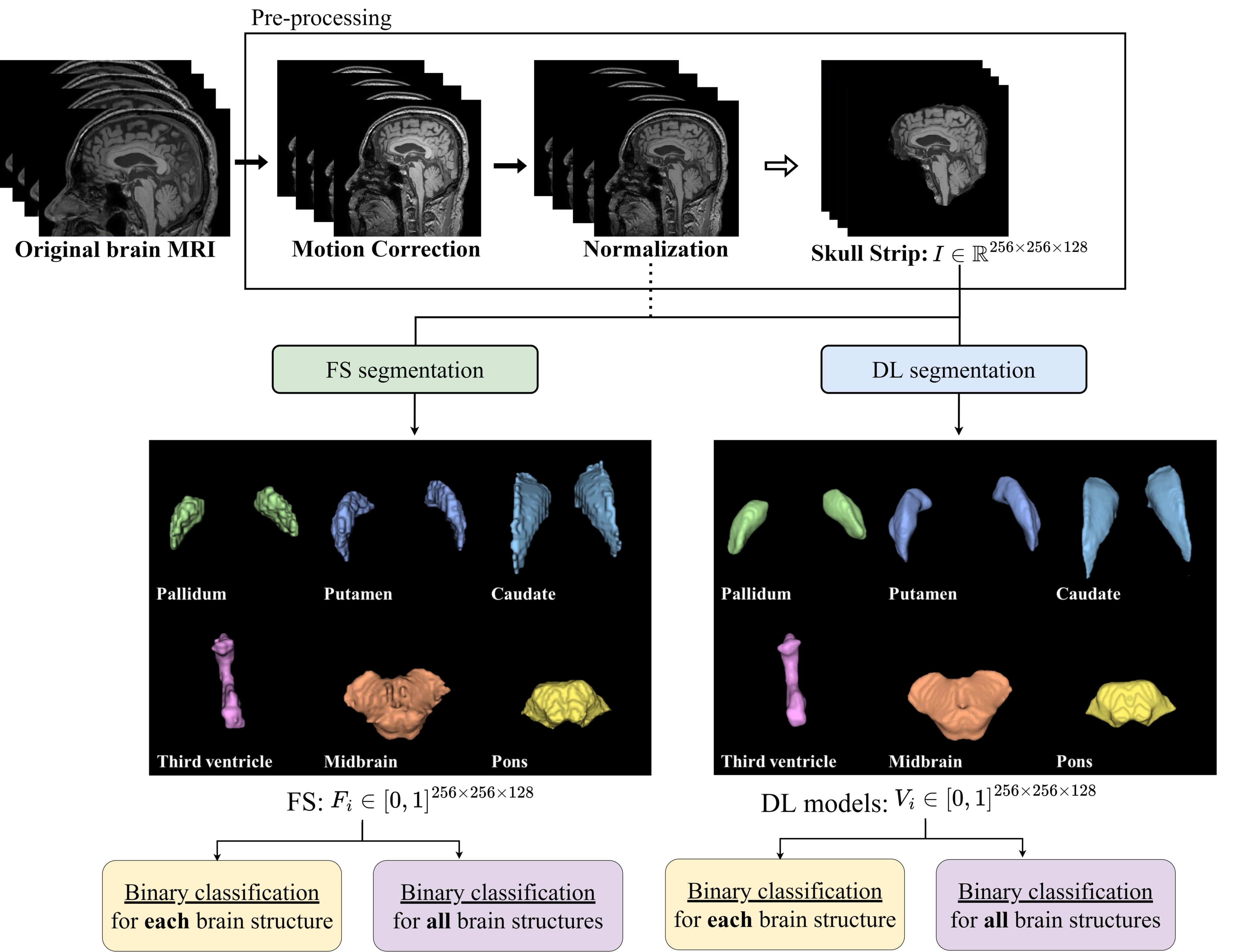}
\caption{Overview of the study. The diagnostic performance of Parkinsonian syndrome regarding analysis time and accuracy for extracting and segmenting brain structures were compared between DL models and FS. Disease diagnosis was performed using the extracted structures individually or comprehensively.}
\label{fig:overview}
\end{figure} 

\section{Methods}

In this section, we describe the brain MRI data (Section \ref{sec:data}), FS implementation (Section \ref{sec:fsmethod}), and DL method implementation (Section \ref{sec:dlmethod}) for the volumetric analysis of key brain structures to diagnose neurodegenerative diseases. Figure \ref{fig:overview} shows an overview of the study process considering the evaluation and comparisons between FS and DL models (i.e., modified V-Net and UNETR representing CNN and ViT {DL architectures, respectively}). Figure \ref{fig:comparison} shows a diagram of the overall performance comparison. We developed DL models with faster processing but similar segmentation performance to FS. The DL models were trained to reproduce and segment the results of FS for each brain structure $F_i  \in [0,1]^{256 \times 256 \times 128}$ as model output $V_i \in [0,1]^{256 \times 256 \times 128}$ by taking skull-stripped brain image $I \in \mathbb{R}^{256 \times 256 \times 128}$ as input ($i \in \{\textup{pallidum, putamen, caudate, third ventricle, midbrain, pons}\}$), with resolution ($h, w, d$) (height $h=256$, width $w=256$, depth $d=128$). The DL segmentation results for the six brain structures were stored as 3D binary masks ($F_i$ and $V_i$ indicate the FS and DL-model masks for brain structure $i$, respectively), where each mask output contained intensities between 0 and 1 (area outside and inside the target brain structure, respectively). By calculating the absolute volume of each or all the brain structures predicted by FS or DL models, we performed binary classification of PD, MSA-C, MSA-P, PSP, and normal cases, and calculated the area under the curve (AUC) of segmentation. 

\subsection{Data preparation}
\label{sec:data}
\subsubsection{Study population and clinical assessments}

\begin{table*}[h]
\caption{Demographic and clinical characteristics of patients enrolled in this study. Data are shown as mean $\pm$ standard deviation or $n$ (\%). PD, Parkinson’s disease; PSP, progressive supranuclear palsy, MSA-P, multiple systemic atrophy—Parkinsonian type; MSA-C, multiple systemic atrophy—cerebellar type}
\resizebox{\linewidth}{!}{%
\begin{tabular}{cccccc}
\toprule
 & PD ($n=105$) & PSP ($n=69$) & MSA-P ($n=63$) & MSA-C ($n=69$) & Normal ($n=105$)\\
\midrule
Age (years) & $69.83\pm10.14$ & $73.86\pm7.85$ & $71.58\pm9.30$ & $64.6\pm9.04$ & $68.29\pm9.69$ \\
Sex (male) & $56\;(53.33)$ & $45\;(65.22)$ & $46\;(66.67)$ & $31\;(49.21)$ & $52\;(49.52)$  \\
Onset to MRI (years) & $6.02\pm6.09$ & $4.68\pm3.34$ & $4.27\pm2.82$ & $3.01\pm2.63$ &  --  \\
\bottomrule
\end{tabular}
}
\label{tab:dataTable}
\end{table*}

This study was approved by the Institutional Review Board of Samsung Medical Center, and the requirement for written informed consent was waived (approval number: 2021-07-026). We retrospectively screened patients from the Neurology Department of Samsung Medical Center between January 2017 and December 2020. Patients diagnosed with PD, probable MSA, or probable PSP were included in this study. The diagnosis for each patient was determined by movement disorder specialists based on the following criteria: PD was determined according to the United Kingdom PD Society Brain Bank criteria \cite{hughes1992accuracy} using [18F] N-(3-fluoropropyl)-2$\beta$-carbon ethoxy-3$\beta$-(4-iodophenyl) nortropane positron emission tomography, while probable MSA and PSP were diagnosed according to the second consensus diagnosis of MSA \cite{gilman2008second} and movement disorder society clinical diagnostic criteria for PSP \cite{hoglinger2017clinical}, respectively. MSA cases were further classified as either MSA-P or MSA-C after reaching consensus \cite{gilman2008second}. Patients with concomitant or structural brain lesions, including stroke and tumors, which may affect brain MRI scans, were excluded from the study. An age-matched healthy elderly population was included as the control group.
Demographic information on age, sex, and disease duration until the brain MRI examination was collected, as listed in Table \ref{tab:dataTable}. We analyzed the data from 411 individuals and performed threefold cross-validation to train and evaluate the DL models. Each group consisted of 105 healthy controls and 105 PD, 69 PSP, 69 MSA-C, and 63 MSA-P cases. 

We applied cross-validation with three outer folds for evaluation to mitigate bias in the validation and test sets and analyze the effect of set composition (combinations of cases in groups). The data were randomly divided into three sections, one for testing and two for training. Each group comprised 35 normal, 35 PD, 23 PSP, 23 MSA-C, and 21 MSA-P cases.

\subsubsection{Data acquisition and standardization}

Axial brain MRI scans were acquired using a standard protocol for T1-magnetization-prepared rapid acquisition of gradient echo, with repetition/echo time of 11,000/125 ms, inversion time of 2,800 ms, field of view of 240 mm, acquisition matrix size of $320 \times 249$, echo train length of 27, 1 signal average, slice thickness of 5 mm, interslice gap of 1.5 mm, and scanning time of 198 s.

We included six brain structures that are involved in Parkinsonian syndromes in the gray matter, namely, the midbrain, pons, putamen, pallidum, caudate, and third ventricle. These areas are reported to have the highest sensitivity and specificity for differentiating Parkinsonian syndromes \cite{saeed2020neuroimaging, fawzi2021brain}. The MRI scans were resized to $256 \times 256 \times 128$ (i.e., number of slices in the coronal/sagittal/axial planes) to segment each structure.

The FS accepts Digital Imaging and Communications in Medicine (DICOM) or Neuroimaging Informatics Technology Initiative (NIfTI) files as inputs. DICOM is a compelling and flexible but complex format that provides interoperability between several hardware and software tools. Given its complexity, DICOM may be inefficient in image processing and analysis \cite{JSSv044i06}. In addition, DICOM stores a single volume as a series of 2D slices, which is cumbersome for 3D imaging. NIfTI is a more straightforward format than DICOM and preserves the essential metadata. In addition, it maintains the volume as a single file and uses raw data after a simple header, and NIfTI files can be loaded and processed faster than DICOM files. Therefore, we converted files in the brain MRI DICOM format into files in the NIfTI format using MRIcroGL\footnote{https://www.nitrc.org/projects/mricrogl/}. 

\subsection{Brain structure segmentation: Baseline with FS}
\label{sec:fsmethod}
The extraction of brain structures obtained using atlas-based automated segmentation are necessary for training and validation before establishing an automated DL segmentation model. In this study, we used these results as DL ground-truth labels and evaluated the validity of DL model for generating the same label. As a representative technology for atlas-based automated segmentation (see details in Supplementary Section \ref{related work}), we selected FS (version 7.2), which is publicly available for neuroscience research and provides high segmentation performance \cite{dewey_hana_2010, eggert_sommer_2012, mayer_latal_2016, klauschen_goldman_2009, heinen_bouvy_2016, velasco_2017}.

{To segment and extract the six brain structures using FS, it sequentially executes the recon-all pipeline\footnote{https://surfer.nmr.mgh.harvard.edu/fswiki/recon-all} and brainstem substructure pipeline\footnote{https://surfer.nmr.mgh.harvard.edu/fswiki/BrainstemSubstructures}. We used both pipelines because the recon-all pipeline does not support segmentation of brainstem structures (e.g., pons and midbrain).  However, because the brainstem substructure pipeline receives preprocessed inputs from the recon-all pipeline, both pipelines should be executed. Therefore, the extraction of the six brain structures through FS can be divided into MRI scan preprocessing in the recon-all pipeline and the remaining segmentation of the recon-all pipeline along with segmentation in the brainstem substructure pipeline. These processes are explained in Sections \ref{sec:fs_mri_pre} and \ref{sec:fs_seg}.}

\subsubsection{MRI scan preprocessing for FS: Motion correction and skull removal}
\label{sec:fs_mri_pre}
The MRI scan preprocessing in the recon-all pipeline of FS mainly consists of 1) motion correction, 2) normalization, and 3) skull stripping. Motion correction is conducted before averaging when various source volumes are used, compensating for small motion variations between volumes. FS constructs cortical surface models and the boundary between white matter and cortical gray matter to automatically match the brain images of patients, using software \cite{fischl2012freesurfer}. In addition, intensity normalization is applied to the original volume. However, adjusting for intensity fluctuations may hinder intensity-based segmentation. Instead, we scale the intensities of all voxels to the mean value (110) of white matter. 

After correcting for motions and normalizing the data, FS removes the skull and provides the {skull-stripped} brain MRI scan. Removing intracranial brain cavities (e.g., skin, fat, muscle, neck, and eyeballs) may reduce human rater variability \cite{KLEESIEK2016460} and promote automated brain image segmentation and improve analysis quality. Therefore, brain MRI scans should be preprocessed to isolate the brain from extracranial or nonbrain tissues in a process known as skull stripping \cite{kalavathi_prasath_2015}. FS developers devised and applied in-house automated skull-stripping algorithms to isolate intracranial cavities by default. 

In this study, the steps of brain MRI scan preprocessing (i.e., skull stripping with motion correction and normalization of a brain MRI scan) took approximately 20 min. We converted the final skull-stripped images to NIfTI files with size of $256 \times 256 \times 128$, while the original brain MRI scan had a size of $256\times 256 \times 256$, which was adjusted for efficient comparison with the DL models. 

\bigskip

\subsubsection{FS for brain structure segmentation}
\label{sec:fs_seg}

After preprocessing (Section \ref{sec:fs_mri_pre}), FS segments the six brain structures by applying the remaining processes of the recon-all pipeline and the complete brainstem substructure pipeline. {After skull stripping, registration-based segmentation proceeds as follows. FS determines and refines the white and gray matter interfaces for both hemispheres. Then, FS searches for the edge of the gray matter, which represents the pial surface.  With pial surfaces, FS expands and inflates sulci banks and gyri ridges. Subsequently, it extends again into a sphere and parcellates the cortex. After applying these processes, FS segments the brain. The recon-all pipeline encompasses some brain structures (i.e., putamen, caudate, pallidum, and third ventricle), while the brainstem substructure pipeline segments the midbrain and pons. }

In this study, the final segmentation result was assessed with the same input size of $256 \times 256 \times 128$\footnote{ The original size of the segmentation result was $256 \times 256 \times 256$, but it was adjusted to $256 \times 256 \times 128$ for comparison with the DL models.}. In addition, we replaced FS with a DL model applied to the skull-stripped MRI scan (i.e., preprocessing result of the recon-all pipeline) to perform segmentation. For the replacement, we evaluated whether the DL analysis is faster than FS analysis and whether the segmentation result of DL is sufficiently reproducible compared with that of FS. The difference between FS and DL segmentation is illustrated in Figures  \ref{fig:brainparts} and  \ref{fig:brainparts_UNETR}.

\subsection{DL models for brain structure segmentation}
\label{sec:dlmethod}

In this study, we used DL models and FS to segment the same skull-stripped images (i.e., images preprocessed by the FS recon-all pipeline, as described in Section \ref{sec:fs_mri_pre}). The original size of the skull-stripped image generated by FS was $256 \times  256 \times  256$, which was adjusted to $256 \times  256 \times 128$ for DL segmentation owing to the limited graphics processing unit (GPU) memory. Specifically, similar to the segmentation using FS described in Section \ref{sec:fs_seg}, the DL models received skull-stripped images as inputs and were trained to individually segment each structure as a binary mask, in which pixels inside and outside the structure were valued 1 and 0, respectively. We evaluated and compared the performance and analysis time of the DL models by replacing the segmentation process of FS after skull stripping with DL. FS may be inefficient because it segments the entire brain image, requiring many hours of processing. In fact, FS takes at least 4.5 h to segment the six brain structures considered in this study because it requires atlas-based registration to transform the coordinates of the entire MRI scan to segment specific brain structures. Consequently, FS cannot notably reduce the processing time even if only six brain structures were to be segmented. On the other hand, we verified that DL segmentation (e.g., using V-Net or UNETR) takes less than 1 min per case to segment the six target brain structures. As DL models do not require complex registration, unlike non-artificial-intelligence methods (e.g., FS), they can substantially increase the processing efficiency.

\subsubsection{DL models}
The implementation details of the DL models are described herein. As DL models, we adopted the CNN-based V-Net \cite{milletari2016vnet} and ViT-based UNETR \cite{hatamizadeh2022unetr} using the segmentation results provided by FS as labels (Section \ref{sec:fs_seg}). The two models were trained to reproduce FS segmentation.

\paragraph{CNN-based V-Net}

\begin{figure}[t]
\centering
\includegraphics[width=\textwidth]{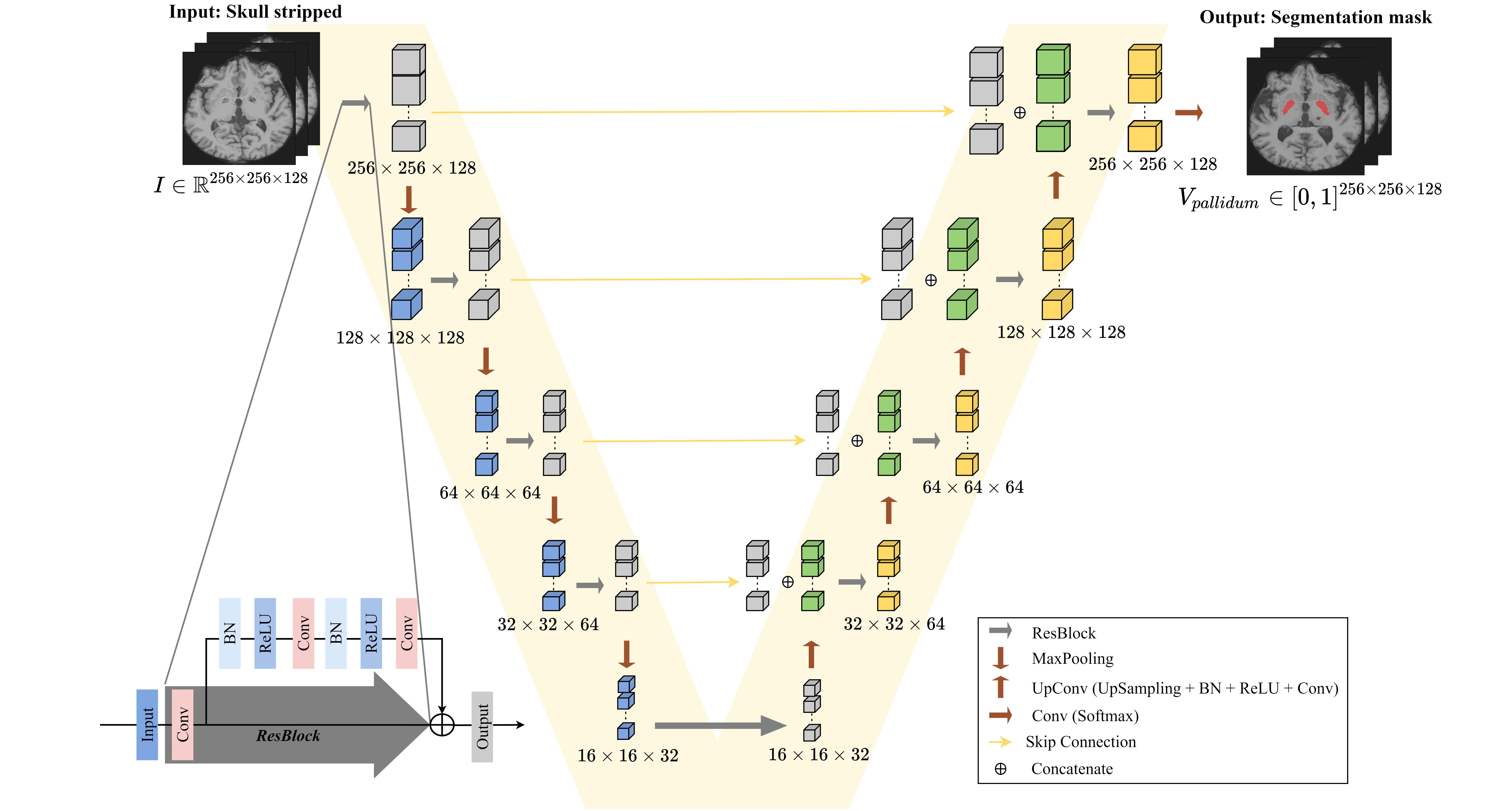}
\caption{Architecture of CNN-based 3D segmentation using V-Net. ResBlock, MaxPooling, and UpConvolution were used to reduce the depth, height, and width. The output shown in the figure is the segmentation of pallidum. (Conv, convolution layer; BN, batch normalization)}
\label{fig:vnet}
\end{figure}

V-Net has been used to segment an entire volume after training an end-to-end CNN on MRI volumes for revealing the prostate \cite{milletari2016vnet}. The architecture of V-Net is V-shaped, where the left part of the network is a compression path, whereas the right part decompresses the features until the original input size is recovered. The left part of the network is separated into stages that operate at varying resolutions. 

In this study, one to three convolutional layers were used in each step. A residual function was learned at each level. The input of the residual part was used in the convolutional layers and nonlinear operations. This output was added to the last convolutional layer of the stage. The rectified linear unit (ReLU) was used as the nonlinear activation function. Convolutions were applied throughout the compression path. The right part of the network learned a residual function similar to that of the left part.
V-Net has shown promising segmentation results, and using this model in our application improved performance. The model was adjusted according to the available memory. The proposed architecture is illustrated in Figure \ref{fig:vnet}. The left part used a residual block (ResBlock) and maximum pooling (MaxPooling). ResBlock was applied to all the blocks with an input size of $256 \times 256 \times 128$. On the other hand, 3D MaxPooling reduced the depth, height, and width of the feature maps to reduce their resolution. The right part also used ResBlock but replaced MaxPooling with UpConvolution, which consisted of 3D upsampling, batch normalization, ReLU activation, and convolutional layers ($5 \times 5\times 5$ filter, same padding, and stride of 1). Upsampling increased the resolution of the feature maps, and batch normalization improved convergence throughout the network \cite{ioffe2015batch}.

\paragraph{ViT-based UNETR}

\begin{figure}[h]
\centering
\includegraphics[width=0.9\textwidth]{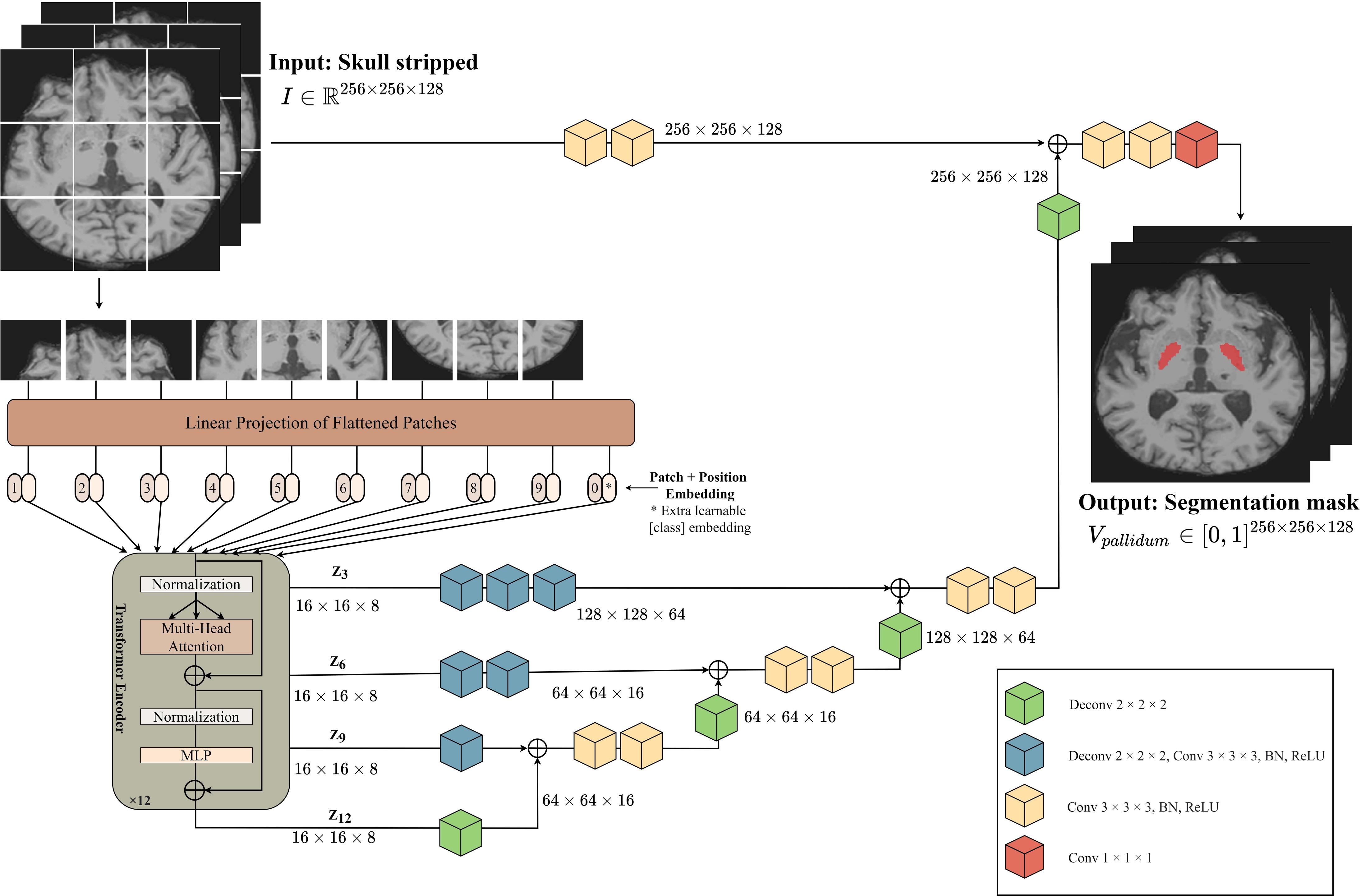}
\caption{Architecture of ViT-based UNETR directly connected to a CNN-based decoder via skip connections at different resolutions for segmentation. (Deconv, deconvolution layer; Conv, convolution layer; BN, batch normalization; MLP, multilayer perceptron)}
\label{fig:unetr}
\end{figure}

UNETR \cite{hatamizadeh2022unetr} is a transformer architecture for 3D medical-image segmentation. It uses a transformer as the encoder to learn the sequence representations of the input volume and capture global multiscale information while adopting U-shaped architectures for the encoder and decoder.
The proposed architecture is illustrated in Figure \ref{fig:unetr}. UNETR followed a contracting--expanding path with an encoder comprising a stack of transformers connected to a decoder through skip connections. The encoder directly used 3D patches and was connected to a CNN-based decoder via a skip connection. A 3D input volume was split into homogeneous nonoverlapping patches and projected onto a subspace using a linear layer. Position embedding was applied to the sequence and then used as input to the transformer. The encoded representations at different levels in the transformer were retrieved and sent to a decoder via skip connections to obtain the segmentation results.

\subsubsection{Implementation details of DL models: Training and inference} 
For the DL models, the input comprised a brain mask and the corresponding patient’s segmented brain structures in the MRI scans, which were merged into an array of dimension $256 \times 256 \times 128$. The ground truth of each brain structure was segmented using FS. For evaluation, threefold cross-validation of the test data was applied to calculate the Dice score and Dice loss. We implemented V-Net in TensorFlow\footnote{https://www.tensorflow.org} and Keras\footnote{https://keras.io} and trained it for 100 epochs.  For UNETR, PyTorch\footnote{https://pytorch.org/} and MONAI\footnote{https://monai.io/} were applied, and the model was trained for 20,000 iterations. Both models were trained using an NVIDIA Tesla V100 DGXS GPU with a batch size of 1 and an initial learning rate of 0.0001.

We evaluated the accuracy of the evaluated models using the Dice score by comparing the expected segmentation with V-Net (or UNETR) and FS outputs. The Dice score measures the overlap between the reference and predicted segmentation masks. A Dice score of 1 indicates perfect spatial correspondence between the two binary pictures, whereas a score of 0 indicates no correlation. We used the Dice loss to determine the performance of the three outer cross-validations on their test sets for the corresponding structures. If $F_i$ and $V_i$ are the ground-truth mask and its prediction for each brain structure, respectively (i.e., FS segmentation mask $F_i$ and its DL prediction mask $V_i$, respectively, as shown in Figure \ref{fig:overview}), the Dice score \cite{sheller_2020} for each brain structure $i \in$ $\{$pallidum, putamen, caudate, third ventricle, midbrain, pons$\}$ is derived as
\begin{ceqn}
\begin{equation}
Dice = \frac{2||{V_i}\circ{F_i}||_{1}}{||{V_i}||_{1}+||{F_i}||_{1}},
\end{equation}
\end{ceqn}
where $\circ{}$ denotes the Hadamard product (i.e., component-wise multiplication) and $||\cdot||_{1}$ is the L1-norm (i.e., sum of absolute values of all components). 
Moreover, we measured the segmentation time for evaluation.

\subsubsection{Statistical analysis for binary classification of cases} 
We obtained the absolute volumes from the six segmented brain structures (i.e., pons, putamen, pallidum, midbrain, caudate, and third ventricle) predicted by the DL models (i.e., CNN-based V-Net or ViT-based UNETR) or FS. Based on the absolute volume of the individual brain structures, we calculated the AUC of the binary classification of diseases, normal vs. P-plus, normal vs. PD, and PD vs. P-plus cases. The AUC was computed based on the receiver operating characteristic curve produced by the correlation between the predicted absolute volume of each brain structure and each case.

Disease binary classification was conducted using the six segmented brain structures individually or collectively. For individual analysis, the AUC was derived through thresholding-based binary classification by obtaining the absolute volume of the individual structures. For a comprehensive analysis of all structures, we additionally considered an ML classification algorithm to perform disease binary classification with the six volumes as inputs. For the classification algorithm, binomial logistic regression (LR) and extreme gradient boosting (XGBoost) were used. LR is a statistical model widely used in ML classification \cite{Austin_2013, thabtah_2019, NUSINOVICI202056}. XGBoost is a well-established method that produces advanced results among gradient-boosting-based techniques \cite{Friedman_2001} (e.g., XGBoost successfully won 17 out of the 29 ML tasks posted on Kaggle by 2015 \cite{Adeola_2020}). In both methods, we evaluated the AUC obtained by the DL model and FS through threefold cross-validation.

\bigskip

\section{Results}

\subsection{Segmentation time of brain structures}

\begin{table*}[h]
\caption{Measured segmentation time per patient obtained by using CNN-based V-Net, ViT-based UNETR, and FS. The time was calculated after the skull-stripped image was obtained. Data are shown as mean $\pm$ standard deviation. (V3, third ventricle)}
\label{tab:TIMETABLE}
\centering
\resizebox{0.65\columnwidth }{!}{%
\begin{tabular}{lcc|c}
\toprule
         & CNN (s)    & ViT (s)  & FS (s) \\
\midrule
Midbrain &  $0.5827\pm0.17$   & $7.5817\pm0.47$  &  \multirow{2}{*}{ $1,698\pm0.144$} \\
Pons     &  $0.5803\pm0.16$   & $9.2242\pm2.02$  &            \\ \hline 
V3       &  $0.5800\pm0.16$   & $7.7525\pm0.45$  &  \multirow{4}{*}{$14,037\pm1.5$} \\
Caudate  &  $0.5749\pm0.16$   & $7.6610\pm0.23$  &            \\
Putamen  &  $0.5815\pm0.17$   & $7.8112\pm0.47$  &            \\
Pallidum &  $0.5847\pm0.18$   & $8.1019\pm0.93$  &            \\ \hline
Total    &  $\mathbf{3.48\pm0.17}$     & $\mathbf{48.14\pm0.97}$  &  $\mathbf{15,735\pm1.07}$          \\
\bottomrule           
\end{tabular}%
}
\end{table*}

\begin{figure}[h]
\centering
\includegraphics[width=0.8\textwidth, height = 0.6\textheight]{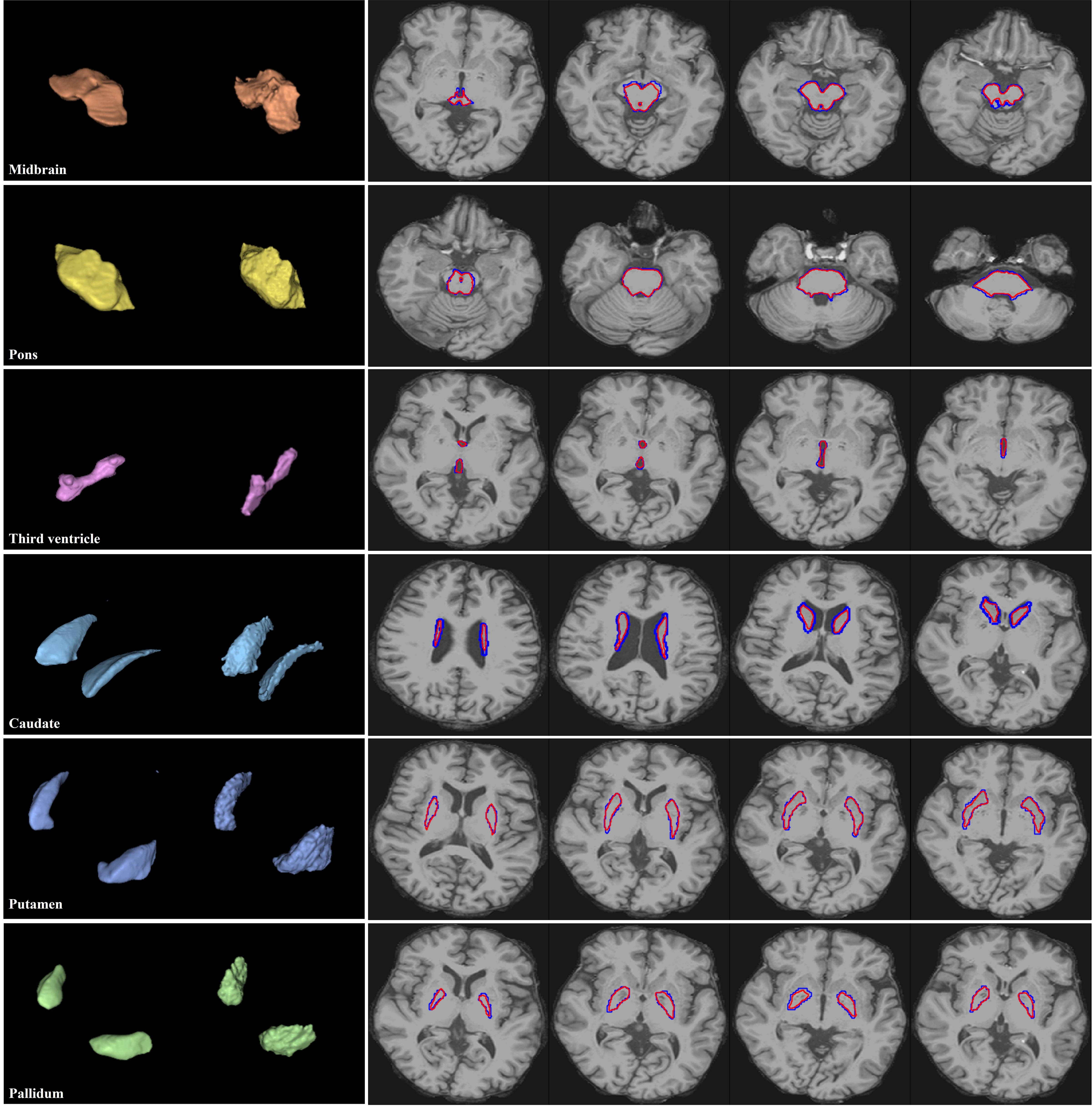}
\caption{Segmentation results of CNN-based V-Net (left 3D images in first column and red-highlighted areas in second column) and FS (right 3D images in first column and blue-highlighted areas in second column) for each brain structure.}
\label{fig:brainparts}
\end{figure}

\begin{figure}[h]
\centering
\includegraphics[width=0.8\textwidth, height = 0.6\textheight]{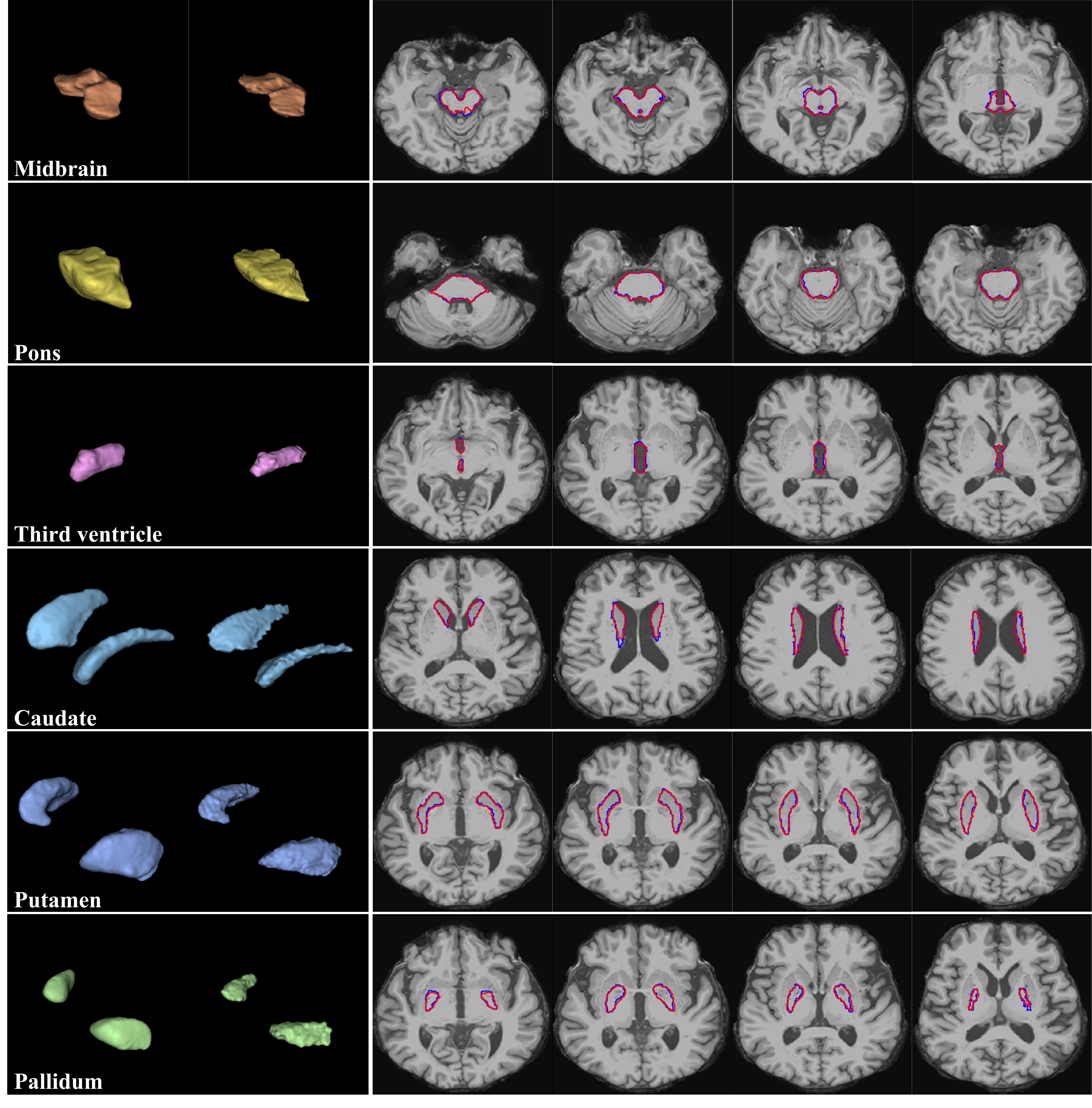}
\caption{Segmentation results of ViT-based UNETR (left 3D images in first column and red-highlighted areas in second column) and FS (right 3D images in first column and blue-highlighted areas in second column) for each brain structure.}
\label{fig:brainparts_UNETR}
\end{figure}

Table \ref{tab:TIMETABLE} lists the time required to segment the six brain structures per patient. As mentioned in Section \ref{sec:fs_seg}, the brain structure segmentation using FS sequentially processes the remaining of the recon-all pipeline and the complete brainstem substructure pipeline. In FS segmentation, we removed the analysis time of preprocessing (i.e., time to extract the skull-stripped image from the original MRI) described in Section \ref{sec:fs_seg}. The resulting time provides a fair comparison of the total times, as FS and DL models use the skull-striped MRI scan as input to derive the final segmentation results, indicated by bold values in Table \ref{tab:TIMETABLE}.

The CNN-based V-Net and ViT-based UNETR are considerably faster than FS. On average, V-Net took 3.48 s to segment the six brain structures, and UNETR took 48.14 s, whereas FS took approximately 15,735 s, being approximately 4521 and 326 times slower than V-Net and UNETR, respectively.

\subsection{Dice score of brain structure segmentation using DL models}

\begin{table*}[h]
\caption{Dice scores of CNN-based V-Net and ViT-based UNETR for brain structures: midbrain, pons, third ventricle (V3), caudate, putamen, and pallidum. Mean $\pm$ standard deviation for threefold cross-validation are provided. }
\centering
\resizebox{0.5\columnwidth}{!}{%
\begin{tabular}{lc|c}
\toprule
         & CNN                 & ViT      \\
\midrule
Midbrain & $0.9385\pm0.020$  &  $0.9642\pm0.001$  \\
Pons     & $0.9661\pm0.010$  &  $0.9748\pm0.006$  \\
V3       & $0.9254\pm0.036$  &  $0.9550\pm0.001$  \\
Caudate  & $0.8892\pm0.038$  &  $0.9456\pm0.001$  \\
Putamen  & $0.8917\pm0.028$  &  $0.9474\pm0.002$  \\
Pallidum & $0.8495\pm0.045$  &  $0.9274\pm0.002$  \\
\bottomrule
\end{tabular}%
}
\label{tab:dicescore}
\end{table*}

Segmentation and prediction results of V-Net and FS are illustrated in Figure \ref{fig:brainparts}. The corresponding results of UNETR are illustrated in Figure \ref{fig:brainparts_UNETR}. The Dice score was obtained (Table \ref{tab:dicescore}) to evaluate the performance of 3D image segmentation. The CNN- and ViT-based models showed high Dice scores above 0.85 for all the brain structures. The Dice scores were higher for the midbrain and pons than for the basal ganglia (i.e., caudate, putamen, pallidum), possibly because the brainstems are surrounded by cerebrospinal fluid and provide a stronger contrast for accurate segmentation. 
The ViT-based model showed a higher Dice score than the CNN-based model, which in turn showed a much shorter segmentation time than the ViT-based model (e.g., 3.48 s for V-Net and 48.14 s for UNETR, as shown in Table \ref{tab:TIMETABLE})\footnote{Although we evaluated V-Net and UNETR in different development environments of TensorFlow and PyTorch, respectively, we expect the CNN-based V-Net to be competitive in speed with the ViT-based UNETR given the segmentation speed difference of at least 10 times in our experiments.} In addition, the CNN-based V-Net had a similar performance to the ViT-based UNETR in actual disease classification, as listed in Table \ref{tab:AUCtable}.

\begin{table*}[h]
\caption{Disease binary classification based on individual brain structures. Segmentation AUC of CNN-based V-Net, ViT-based UNETR, and FS. Mean $\pm$ standard deviation for threefold cross-validation and midbrain-to-pons ratio segmentation are listed.} 
\resizebox{\columnwidth}{!}{%
\begin{tabular}{llccccccc}
\toprule
Case &  & Midbrain & Pons & Midbrain/pons & V3 & Caudate & Putamen & Pallidum \\
\midrule
\multirow{2}{*}{Normal vs. PSP} & V-Net & ${0.73\pm0.06}$ & ${0.69\pm0.03}^*$ & ${0.65\pm0.08}$ & ${0.83\pm0.09}^*$ & ${0.54\pm0.04}$ & ${0.74\pm0.01}^*$ & ${0.78\pm0.05}^*$ \\
& UNETR & ${0.69\pm0.06}$ & ${0.64\pm0.05}^*$ & $0.60\pm0.07$ & ${0.84\pm0.08}^*$ & ${0.57\pm0.03}$ & ${0.69\pm0.03}^*$ & ${0.76\pm0.05}$ \\
& FS & $0.70\pm0.06$ & $\mathbf{0.89\pm0.05}$ & $0.65\pm0.08$ & $0.82\pm0.02$ & $0.56\pm0.06$ & $0.62\pm0.09$ & $0.72\pm0.11$  \\
\\
\multirow{2}{*}{Normal vs. MSA-P} & V-Net & ${0.63\pm0.05}$ & ${0.60\pm0.05}$ & $\mathbf{0.73\pm0.03}$ & $\mathbf{0.73\pm0.03}$ & ${0.61\pm0.02}$ & $\mathbf{0.73\pm0.03}$ & ${0.67\pm0.10}$  \\
& UNETR & ${0.61\pm0.05}$ & ${0.67\pm0.06}$ & ${0.70\pm0.05}$ & ${0.70\pm0.03}$ & ${0.59\pm0.03}$ & $\mathbf{0.73\pm0.03}$ & $0.65\pm0.09$\\
& FS & $0.64\pm0.04$ & $0.65\pm0.04$ & $0.70\pm0.05$ & $\mathbf{0.73\pm0.03}$ & $0.60\pm0.06$ & $0.70\pm0.03$ & $0.66\pm0.08$  \\
\\
\multirow{2}{*}{Normal vs. MSA-C} & V-Net & ${0.76\pm0.11}$ & ${0.90\pm0.04}$ & $\mathbf{0.91}\pm\mathbf{0.02}$ & ${0.56\pm0.02}$ & ${0.61\pm0.01}$ & ${0.65\pm0.11}$ & ${0.66\pm0.09}$ \\
& UNETR & ${0.73\pm0.08}$ & ${0.86\pm0.06}$ & ${0.81\pm0.15}$ & ${0.58\pm0.01}^*$ & $0.57\pm0.04$ & $0.62\pm0.10$ & ${0.66\pm0.07}$ \\
& FS & $0.76\pm0.10$ & $0.90\pm0.04$ & $\mathbf{0.91\pm0.02}$ & $0.56\pm0.01$ & $0.62\pm0.10$ & $0.65\pm0.10$ & $0.71\pm0.09$ \\
\\
\multirow{2}{*}{Normal vs. PD} & V-Net & ${0.55\pm0.02}$ & ${0.53\pm0.02}$ & ${0.57\pm0.04}$ & ${0.61\pm0.07}$ & ${0.57\pm0.02}$ & ${0.55\pm0.03}$ & ${0.56\pm0.02}$\\
& UNETR & ${0.58\pm0.03}$ & $0.55\pm0.03$ & $0.53\pm0.04$ & $\mathbf{0.63\pm0.06}$ & $0.55\pm0.04$ & $0.54\pm0.05$ & $0.54\pm0.03$\\
& FS & $0.56\pm0.02$ & $0.52\pm0.01$ & $0.57\pm0.04$ & $0.61\pm0.08$ & $0.54\pm0.02$ & $0.57\pm0.01$ & $0.54\pm0.02$\\
\\
\multirow{2}{*}{PD vs. PSP} & V-Net & ${0.71\pm0.07}$ & ${0.67\pm0.03}$ & ${0.58\pm0.07}$ & ${0.74\pm0.09}$ & ${0.54\pm0.02}$ & ${0.67\pm0.03}$ & $\mathbf{0.75\pm0.03}$\\ 
& UNETR & $0.67\pm0.09$ & $0.63\pm0.07$ & $0.57\pm0.08$ & ${0.72\pm0.10}$ & ${0.52\pm0.01}$ & ${0.63\pm0.01}$ & ${0.72\pm0.05}$ \\
& FS & $0.69\pm0.06$ & $0.65\pm0.04$ & $0.58\pm0.07$ & $0.74\pm0.11$ & $0.52\pm0.01$ & $0.62\pm0.03$ & $0.71\pm0.08$\\
\\
\multirow{2}{*}{PD vs. MSA-P} & V-Net & ${0.59\pm0.07}$ & ${0.67\pm0.02}$ & ${0.58\pm0.06}$ & ${0.65\pm0.08}$ & ${0.58\pm0.02}$ & ${0.67\pm0.01}^*$ & ${0.65\pm0.07}$\\
& UNETR & ${0.56\pm0.02}$ & ${0.64\pm0.02}$ & ${0.67\pm0.04}^*$ & $0.59\pm0.05$ & ${0.61\pm0.03}$ & ${0.66\pm0.01}^*$ & ${0.63\pm0.05}$\\
& FS & $0.59\pm0.03$ & $0.66\pm0.02$ & $0.58\pm0.06$ & $0.65\pm0.08$ & $0.59\pm0.04$ & $\mathbf{0.69\pm0.04}$ & $0.66\pm0.07$\\
\\
\multirow{2}{*}{PD vs. MSA-C} & V-Net & ${0.74\pm0.06}$ & ${0.90\pm0.03}$ & $\mathbf{0.94\pm0.02}$ & ${0.56\pm0.08}$ & ${0.59\pm0.06}$ & ${0.50\pm0.07}$ & ${0.64\pm0.06}$\\ 
& UNETR & ${0.69\pm0.02}$ & ${0.82\pm0.11}$ & ${0.81\pm0.16}$ & $0.56\pm0.08$ & ${0.58\pm0.01}$ & $0.57\pm0.05$ & ${0.62\pm0.05}$\\
& FS & $0.71\pm0.06$ & $0.90\pm0.03$ & $\mathbf{0.94\pm0.02}$ & $0.57\pm0.08$ & $0.59\pm0.06$ & $0.59\pm0.08$ & $0.69\pm0.10$\\
\\

\bottomrule
\multicolumn{9}{l}{\small $^*$ $p < 0.05$ indicates a significant difference in AUC between the DL models and FS.} \\
\multicolumn{9}{l}{\small {The best result for each volume segmentation method based on FS and DL in binary classification is shown in bold.}}
\end{tabular}%
}
\label{tab:AUCtable}
\end{table*}

\subsection{Binary classification based on individual brain structures}

Using the estimated volumes, we performed binary classification for cases normal vs. P-plus, normal vs. PD, and PD vs. P-plus, where P-plus comprised PSP, MSA-P, and MSA-C cases. The AUCs of the brain structures for each model were compared, as listed in Table \ref{tab:AUCtable}, which also presents the AUC of the midbrain-to-pons ratio \cite{chougar2021automated}. 

Among the 98 cases (7 cases of binary classification × 2 DL models × 7 cases of brain structures), there was no significant difference in AUC between the DL models and FS, except for 11 cases. Of these 11 cases, 7 AUCs of the DL models (i.e., CNN-based V-Net and ViT-based UNETR) were higher than those of FS. Furthermore, most of the cases for the CNN-based V-Net showed no lower AUC for disease classification than the cases for the ViT-based UNETR.

The highest AUCs in the comparison between the methods were higher in normal or PD vs. MSA-C (0.91––0.94) than in normal or PD vs. PSP (0.75-–0.89). Among the brain structures, the midbrain-to-pons ratio showed the best performance in normal vs. MSA-C and PD vs. MSA-C, while the third ventricle and pallidum showed the best performance in normal vs. PSP and PD vs. PSP. The highest AUCs were not significantly different in the classification of normal or PD vs. MSA-P (0.69–-0.73) or PD (0.63).

\subsection{Binary classification based on complete brain structures}

Most AUCs of the DL models were not significantly different from those of FS, as listed in Table \ref{tab:AUCregression}, although a considerable difference existed in the segmentation speed between the models and FS, as listed in Table \ref{tab:TIMETABLE}.
In Table \ref{tab:AUCregression}, the highest AUC of FS and DL models for each binary classification are indicated in bold. The highest AUCs of classification between PD vs. P-plus and normal vs. P-plus were higher than 0.8 in both DL models, except for PD vs. MSA-P (AUC > 0.76). There was no significant difference between FS and the DL models (p-value of 0.05 or higher) in all highest AUCs.

Table \ref{tab:AUCregression} shows that of the 28 cases (2 ML models $\times$ 2 DL models $\times$ 7 binary classifications), 24 cases had no significant differences with FS, obtaining p-values above 0.05. Like listed in Table 3, 
the CNN-based V-Net achieved a better AUC than the ViT-based UNETR. In 9 of the 14 pairs of cases, the CNN-based V-Net outperformed the ViT-based UNETR.
In both LR and XGBoost, collectively considering the six brain structures (Table \ref{tab:AUCregression})
resulted in a significantly higher AUC than when considering the individual structures (Table \ref{tab:AUCtable}). 
The best performance was higher in normal or PD vs. MSA-C (0.93-–0.95) than in normal or PD vs. PSP (0.80-–0.89). Unlike the AUC in individual brain structures, the highest AUC became significant in normal or PD vs. MSA-P (0.79-–0.82). The highest AUC was not significantly different for normal vs. PD (0.70). We interpret these results in the Discussion section.

\bigskip

\begin{table}[h]
\caption{ Binary classification of diseases based on all the brain structures. AUC in LR and XGBoost of CNN-based V-Net, ViT-based UNETR, and FS. The AUC is expressed as the mean from threefold cross-validation. LR; logistic regression, XGBoost; eXtreme Gradient Boosting}
\label{tab:AUCregression}
\centering
\resizebox{\columnwidth}{!}{%
\begin{tabular}{lcc|cc|cc}
\toprule
Case         & \multicolumn{2}{c|}{V-Net}    & \multicolumn{2}{c|}{UNETR}    & \multicolumn{2}{c}{FS}        \\ \midrule
                & LR & XGBoost & LR & XGBoost & LR & XGBoost \\ \midrule

Normal vs. PSP   & $\mathbf{0.89\pm0.07}$ & ${0.86\pm0.05}$    & ${0.89\pm0.08}$ & ${0.84\pm0.04}$ & $\mathbf{0.89\pm0.07}$  & ${0.87\pm0.06}$  \\
Normal vs. MSA-P & ${0.78\pm0.04}$ & ${0.73\pm0.003}^*$ & $\mathbf{0.81\pm0.03}$ & ${0.77\pm0.04}$ & ${0.79\pm0.001}$ & $\mathbf{0.82\pm0.01}$  \\
Normal vs. MSA-C & ${0.90\pm0.03}$ & $\mathbf{0.93\pm0.04}$    & ${0.85\pm0.12}$ & ${0.90\pm0.10}$ & ${0.88\pm0.04}$ & $\mathbf{0.95\pm0.03}$ \\
Normal vs. PD    & ${0.60\pm0.07}$ & $\mathbf{0.66\pm0.02}^*$  & ${0.60\pm0.04}$ & ${0.60\pm0.03}$ & ${0.65\pm0.07}$ & $\mathbf{0.70\pm0.05}$ \\
PD vs. PSP       & $\mathbf{0.80\pm0.08}$ & ${0.78\pm0.001}^*$   & ${0.77\pm0.13}$ & ${0.75\pm0.01}$ & $\mathbf{0.77\pm0.10}$ & ${0.76\pm0.03}$ \\
PD vs. MSA-P     & $\mathbf{0.76\pm0.07}$ & ${0.66\pm0.03}$    & ${0.71\pm0.02}^*$ & ${0.68\pm0.07}^*$  &  $\mathbf{0.79\pm0.08}$ & ${0.71\pm0.02}$ \\
PD vs. MSA-C     & $\mathbf{0.91\pm0.04}$ & ${0.87\pm0.03}$    & ${0.80\pm0.19}$   & ${0.80\pm0.12}$  &  ${0.89\pm0.07}$ & $\mathbf{0.91\pm0.05}$  \\
\bottomrule 
\multicolumn{7}{l}{\footnotesize The best result for each volume segmentation method based on FS and DL in each binary classification is shown in bold. }\\
\multicolumn{7}{l}{\footnotesize $^*$ $p < 0.05$ indicates a significant difference in AUC between the DL models and FS. } 
\end{tabular}%
}
\end{table}

\section{Discussion}
\label{sec:disc}
We developed two DL models, V-Net and UNETR, which showed significantly faster brain segmentation than FS and a comparable accuracy. Our DL models shortened the segmentation time by at least 300 times compared with FS. Moreover, they showed robust high performance in differential diagnosis between PD and P-plus cases using the volume of segmented brain structures. The DL models were efficient (i.e., analysis speed at least 300 times faster than FS) and effective (i.e., comparable to FS in Dice score and AUC) in automated brain segmentation and disease diagnosis, even for simultaneous analysis of all brain structures and their individual analyses. Thus, the proposed DL models may promote the application of automated brain segmentation in clinical practice and facilitate efficient and accurate brain research in medicine.

Automated tools have scarcely been adopted for brain segmentation in clinical practice despite their high accuracy in the differential diagnosis of patients with Parkinsonism \cite{saeed2020neuroimaging, fawzi2021brain}. This is mainly attributable to the complicated and time-consuming process of automated brain segmentation compared with physicians' qualitative visual assessment of brain MRI scans. Consequently, automated segmentation models have mainly been used in research settings that require quantitative brain measurements. Nevertheless, their application in clinical settings may increase with our DL models, which have shown much faster segmentation than FS with a similar accuracy. The DL models may contribute to improve the accuracy of clinical diagnosis of PD or P-plus cases by providing precise brain image analysis. In addition, clinical trials that require quantitative brain measurement from a large population may be conveniently conducted using our fast and accurate DL models. In the past, methods for brain image analysis were time- and resource-consuming, even with an automated segmentation tool such as FS.

While V-net and UNETR showed significantly faster segmentation with satisfactory accuracy, the CNN-based V-Net may be more suitable in clinical settings for diagnosis based on volumetry of brain MRI scans. Although the ViT-based UNETR is the most recent DL model and shows a high Dice score, the number of training parameters is approximately 46 times larger than that of V-Net. As the number of calculations increases with the number of trainable parameters, the hardware requirements increase in terms of GPU memory and processing power. Consequently, the ViT-based UNETR may be considerably demanding for training and evaluation, requiring high specification GPU. The CNN-based V-Net showed an AUC generally higher than that of UNETR and lower Dice scores. Until the ViT performance is further improved, the CNN-based V-Net, which uses fewer GPU resources, seems to be the best option for clinical practice.

Regarding the AUC of differential diagnosis for PD and P-plus cases, the CNN- and ViT-based models (V-Net and UNETR, respectively) showed comparable performance to FS. { Since our DL models are at least 300 times faster than FS without sacrificing diagnostic performance, they are superior to FS in terms of clinical efficacy.} In binary classification using individual brain structures, the relative order of the AUC of each brain structure was consistent with previously reported results \cite{massey2012conventional, sjostrom2020automated}. For instance, the pons and midbrain-to-pons ratio showed the highest AUC in classification of normal vs. MSA-C and PD vs. MSA-C cases. The third ventricle and pallidum showed the highest AUC in classification of normal vs. PSP and PD vs. PSP cases. The putamen showed the highest AUC in classification of PD and MSA-P cases. In the classification of PD vs. PSP cases, the third ventricle showed a higher AUC, whereas the midbrain showed a relatively lower AUC. Single measurements of the midbrain have failed to differentiate PSP from PD or MSA \cite{brooks2009proposed,hotter2009potential,oba2005new}, despite classic MRI studies showing atrophic midbrain in PSP  \cite{quattrone2008mr,schrag2000differentiation}. On the other hand, the third ventricle has been shown to be a reliable marker for diagnosing early stage PSP from PD and late-stage PSP \cite{quattrone2021new}, and it has been added to a new version of the magnetic resonance Parkinsonism index \cite{quattrone2018new}.

For binary classification based on the six brain structures, significant improvements in the AUC were achieved in all models. In both DL models, the highest AUC of classification of PD vs. P-plus and normal vs. P-plus cases was above 0.8, except for PD vs. MSA-P cases. The relatively low AUC of classification between PD and MSA-P cases based on brain MRI cases has also been reported in previous studies. \cite{massey2012conventional, sjostrom2020automated}. The limitation of clinical diagnosis may have contributed to the relatively low AUCs in these studies owing to the overlapping manifestations between PD and MSA-P cases. Clinical diagnosis of PSP and MSA-P has been reported to have the most frequent discrepancy from autopsy-proven diagnosis, even when considering diagnostic criteria \cite{rizzo2016accuracy}. No significant difference in brain MRI scans has been found between normal and PD cases, resulting in no significant AUC differences for classification between these cases.

Our study has some limitations. First, the diagnoses of PD, PSP, and MSA-C were not pathologically verified. Instead, movement specialists provided clinical diagnoses based on validated clinical consensus, providing only probable diagnosis. Second, we segmented six brain structures, namely, midbrain, pons, medulla, putamen, pallidum, and third ventricle, but disregarded other brain structures that may reflect different pathologic characteristics between PD and P-plus (e.g., cerebellum, middle cerebellar peduncle). We excluded those structures owing to the low segmentation accuracy achieved by FS. Nevertheless, the differential diagnosis of P-plus using only the brain structures included in this study has been reported as reliable \cite{chougar2021automated}. Third, given memory limitations, we downscaled the output shape from $256 \times 256 \times 256$ to $256 \times 256 \times 128$, which may have caused an information loss. Nevertheless, the Dice scores suggest a negligible impact of information loss, whereas using a downscaled input accelerates training and inference in DL models.

Automated segmentation of brain MRI scans has become an influential method for diagnosing neurodegenerative diseases, including movement disorders. The proposed DL models showed remarkable results for both brain segmentation and the differential diagnosis of PD and P-plus. Using the high-performance CNN- and ViT-based models, we significantly shortened the segmentation time of deep brain structures while obtaining comparable accuracy to the conventional FS segmentation. Despite the superior DL performance, no quantitative results of the comparative analysis and evaluation of the performance of DL have been reported to date for the differential diagnosis of neurodegenerative diseases, including PD and P-plus. To the best of our knowledge, this is the first study to quantitatively establish the significance of DL segmentation and disease classification. We found that the cost-effective CNN-based model achieves satisfactory performance in both segmentation and differential diagnosis compared with the most recent ViT-based model. Our DL models may contribute to the development of patient- and clinician-friendly segmentation methods that enable fast and accurate diagnosis and may provide a meaningful reference for hospitals planning to introduce DL brain segmentation and diagnosis for neurodegenerative diseases.



\section{Data Availability}
The authors declare that the main data supporting the results of this study are available within the paper. The raw datasets from Samsung Medical Center are protected to preserve patient privacy but can be made available upon reasonable request provided that approval is obtained from the corresponding Institutional Review Board. 





\clearpage

\appendix
\renewcommand\thefigure{\thesection.\arabic{figure}}    

\section{Related work}  \label{related work}
\subsection{Manual segmentation} \label{sec:manual}
In manual MRI scan segmentation, human raters (e.g., expert physicians) manually delineate and label regions of interest in the scans \cite{zijdenbos1994morphometric}. 
Although this method is considered as the gold standard, it is cumbersome and has low reproducibility. Manual segmentation of 3D volume scans is generally performed slice-by-slice and typically requires segmentation of 80 slices,  being tedious and time-consuming.
While various brain structures have been used to diagnose central nervous system diseases (e.g., stroke, Alzheimer’s disease), certain brain structures are used for diagnosing atypical Parkinsonism. The putamen, globus pallidus, midbrain, and pons are the main brain structures that show changes in atypical Parkinsonism and are often segmented for diagnosis and differentiation. However, manual segmentation of these structures in a brain MRI scan is time-consuming and strenuous, even for an experienced radiologist or neurologist who can accurately recognize these structures. In addition, manual segmentation is prone to inter- and intra-rater variability \cite{massey2012conventional, schrag2000differentiation, kim2015visual}. Moreover, the segmentation quality depends on rater proficiency, and even experienced specialists may show variability from their previous annotations. Hence, validation by at least two raters is required for the analysis. Given the challenges and problems of manual segmentation, automated methods are preferred for large-scale datasets in clinical trials or when accurate and quantitative analyses of brain MRI scans are required, such as when measuring the volume or intensity of signals in a brain structure. 

\subsection{Automated segmentation: Atlas-based method} \label{sec:atlas-seg}
Automated image segmentation has been dominated by atlas-based methods that formulate segmentation as an image-registration problem \cite{pham2000current}. A labeled image (i.e., an atlas) is transformed (i.e., registered) using a deformation model for mapping onto an unlabeled image (i.e., test scan). The established spatial correspondence is then used to transfer labels from the atlas to the target MRI scan \cite{christensen1997volumetric, collins1995automatic, iosifescu1997automated}. 
Initially, a single atlas delineated by medical experts was used, but segmentation could be highly biased depending on the quality of registration (i.e., similarity between the atlas and scan) \cite{billot2020automated}. 
Subsequently, multiple labeled atlases have been used to mitigate bias and capture wide anatomical variations \cite{billot2020automated}. Accordingly, two strategies have been proposed: 1) multi-atlas and 2) Bayesian segmentation. Multi-atlas segmentation registers atlases individually onto the test scan and applies label fusion (majority voting) to propagate the most frequently selected labels \cite{heckemann2006automatic, klein2005mindboggle}.
Bayesian segmentation uses a single probabilistic atlas that summarizes all atlases \cite{ashburner2005unified, fischl2002whole}. This entails propagating label probabilities (prior) and image voxel intensities (likelihood) to deduce a generative model (posterior probability) using Bayes' rule. This strategy can be adapted to MRI scans \cite{ashburner2005unified, puonti2016fast, van1999automated, wells1996adaptive} and is faster than multi-atlas segmentation because it requires only one computationally intensive registration step per scan. 
Bayesian segmentation is implemented in various tools such as FS \cite{fischl2012freesurfer}, statistical parametric mapping \cite{ashburner2005unified}, and the FMRIB software library (FSL) FMRIB integrated registration and segmentation tool (FIRST) \cite{patenaude2011bayesian}.

\citet{velasco_2017} analyzed various automated segmentation algorithms. For our six target brain structures, \citet{velasco_2017} reported the average specificity, positive predictive value, and Dice score of FS as higher than those of FSL-FIRST.  Additionally, compared with other automated approaches (i.e., statistical parametric mapping and FSL), FS had the highest sensitivity and specificity for brain volume changes in ROC analysis, achieving more consistency, less susceptibility to noise, and better image quality \cite{dewey_hana_2010, eggert_sommer_2012, mayer_latal_2016, klauschen_goldman_2009}. Furthermore, with several segmentation tools introduced for general brain segmentation, FS is frequently used in PD diagnosis \cite{messina_2011, vasconcellos_2018, Henrik_2020, SALSONE20141004}. Therefore, with the extensive and automated analysis of key features in the human brain, FS has been widely recognized as the most representative atlas-based automated segmentation method for brain structure analysis. Thus, it served as reference in our study.

\subsection{Automated segmentation: DL model} 
Modern automated image segmentation relies on DL techniques, with the two most generalized DL models being CNNs and ViTs. As for other computer vision tasks, CNNs are predominant in image segmentation owing to the effectiveness of the convolution operation. Convolution deals with sparse interactions (local connections), weight (parameter) sharing, and translation equivariance, giving CNNs a strong and useful inductive bias (prior knowledge) and allowing them to quickly converge with reduced computational complexity. 
Owing to the effectiveness of the convolution operation, the UNet architecture \cite{ronneberger2015u} has achieved outstanding results in the medical field \cite{heinrich2019obelisk, kamnitsas2016deepmedic, oktay2018attention, shen2017deep, tong20173d, zeng20173d}, being widely used for segmentation. UNet has a U-shaped symmetric encoder--decoder architecture, typically including 1) a convolutional encoder (or downsampling network) to extract relevant features from the inputs at different resolutions, followed by 2) a convolutional decoder (or upsampling network) to synthesize the extracted features as a high-resolution image to obtain pixel- or voxel-wise precision, and 3) a skip connection between layers to recover spatial information lost during downsampling. V-Net \cite{milletari2016vnet} is a representative variant of UNet for 3D medical image segmentation. 

Despite their efficiency, CNNs have a limited ability to learn long-distance dependencies owing to the locality of receptive fields in the convolutional layers \cite{yu2015multi, zhao2017pyramid}. Thus, transformer-based models, which use self-attention mechanisms as core operators, have recently enabled attractive solutions for computer vision tasks.
The key idea of the self-attention mechanism, which has shown great success in natural language processing, is to learn the relative importance (self-alignment) of a single token relative to all other tokens in a sequence \cite{bahdanau2014neural}. In other words, calculating the pairwise interactions between all input units has essentially the same effect as having a global receptive field of long-range dependencies \cite{vaswani2017attention}.
Inspired by this mechanism, ViT \cite{dosovitskiy2020image} was introduced to interpret an image as a sequence of patches, adapting self-attention for computer vision applications. ViT and its variants have demonstrated excellent performance in many computer vision tasks \cite{zhu2020deformable, zheng2021rethinking, kumar2021colorization, chen2021pre, arnab2021vivit}. 
UNETR \cite{hatamizadeh2022unetr} is a representative ViT-based 3D image segmentation model that improves the segmentation performance by reducing the loss of encoding information by converting the encoder of an existing CNN-based segmentation model into a ViT.
However, transformer-based approaches have limitations, such as the need for large amounts of training data owing to the lack of inductive bias and the quadratic computational complexity of self-attention according to the input image size \cite{dosovitskiy2020image}.  

Although the medical community has a great interest in DL models for image segmentation, few studies have been conducted on segmenting the intricate brain structures to diagnose diseases. To the best of our knowledge, no existing study has investigated DL methods for segmenting the biomarkers of Parkinsonian syndromes. \citet{bocchetta_Parkinsonian, MANJON2020102184} used FS as an automated segmentation tool toward diagnosing Parkinsonian syndromes but neglecting DL methods. Similarly, \citet{KHAN2008735} proposed a method for fully automated segmentation of the brain without relying on DL. They introduced a pipeline that uses FS labeling to provide information in a highly nonlinear transformation method (i.e., large deformation diffeomorphic metric mapping).
In this study, we used high-performance DL models based on CNN and ViT to segment brain structures of patients with Parkinsonian syndromes. We established that DL models can yield equal or more effective results than FS. These models can substantially shorten the segmentation time while retaining the accuracy of non-DL FS segmentation.

\clearpage
\setcounter{figure}{0}    

\section{Additional qualitative results} \label{add_figures}
\begin{figure*}[h]
\centering
\includegraphics[width=0.9\textwidth, height = 0.7\textheight]{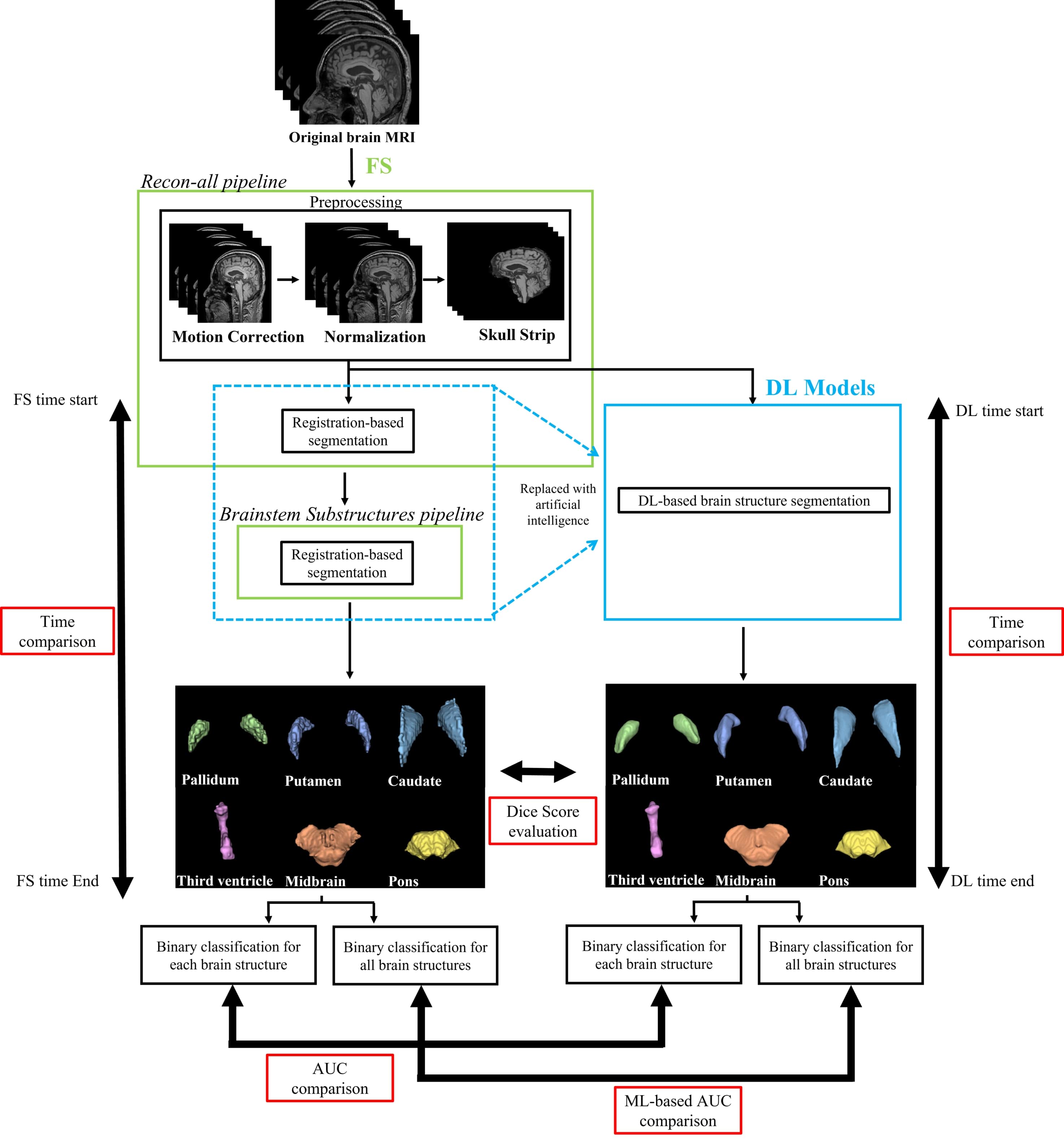}
\caption{Study overview and performance comparisons. Using FS and DL V-Net and UNETR for segmentation, we analyzed the segmentation time, Dice score, and AUC of disease diagnosis considering each brain structure. In addition, a comparison of AUC was conducted using ML methods for disease diagnosis considering all brain structures.}
\label{fig:comparison}
\end{figure*}

\clearpage

\bibliographystyle{unsrtnat}
\bibliography{reference_arxiv.bib}
\clearpage 

\end{document}